\documentclass[smallextended]{svjour3}     % onecolumn (second format)
\smartqed  % flush right qed marks, e.g. at end of proof
\usepackage{graphicx}
\newcommand{\SPACE}{{\sl SPACE}}
\newcommand{\Planck}{{\sl Planck}}
\begin{document}

\title{{\sl SPACE}: the SPectroscopic All-sky Cosmic Explorer} 

\author{A. Cimatti$^1$ \and 
M. Robberto$^2$ \and 
C. Baugh$^3$ \and 
S. V. W. Beckwith$^2$ \and 
R. Content$^4$ \and E. Daddi$^5$ \and 
G. De Lucia$^6$ \and B. Garilli$^7$ \and L. Guzzo$^8$ \and G. 
Kauffmann$^6$
% ...
}

\institute{$^1$Universit\`a di Bologna, Dipartimento di Astronomia, Italy
\and $^2$Space Telescope Science Institute, Baltimore, USA \and
$^3$University of Durham, Institute of Computational Cosmology, UK 
\and $^4$University of Durham, Centre for Advanced Instrumentation, UK 
\and $^5$CEA Saclay, France \and $^6$MPA Garching, Germany \and 
$^7$INAF - IASFMI, Italy \and $^8$INAF - Brera, Italy
% ...
}

\date{Received: November 21, 2007 / Accepted: 10 Apr, 2008}

\maketitle

\begin{abstract}
We describe the scientific motivations, the mission concept and the 
instrumentation of \SPACE, a class-M mission proposed for concept study 
at the first call of the ESA {\sl Cosmic-Vision 2015-2025} planning 
cycle. \SPACE~ aims to produce the largest three-dimensional evolutionary 
map of the Universe over the past 10 billion years  by taking 
near-IR spectra and measuring redshifts for more than half a billion 
galaxies at $0<z<2$ down to $AB\sim23$ over  $3\pi$~sr of the sky. 
In addition, \SPACE~ will also target a smaller sky field, performing 
a deep spectroscopic survey of millions of galaxies to $AB\sim26$ and 
at $2<z<10+$. These goals are unreachable with ground-based observations
due to the $\approx$500 times higher sky background (see e.g. 
\cite{Aldering}). To achieve 
the main science objectives, \SPACE~ will use a 1.5m diameter 
Ritchey-Chretien telescope equipped with a set of arrays of Digital 
Micro-mirror Devices (DMDs) covering a total field of view of 0.4 deg$^2$,
and will perform large-multiplexing multi-object spectroscopy (e.g. 
$\approx$6000 targets per pointing) at a spectral resolution of R$\sim$400 
as well as diffraction-limited imaging with continuous coverage from 
0.8$\mu$m to 1.8$\mu$m. Owing to the depth, redshift 
range, volume coverage and quality of its spectra, \SPACE~ 
will reveal with unique sensitivity most of the fundamental cosmological
signatures, including the power spectrum of density fluctuations and its 
turnover. \SPACE~ will also place high accuracy constraints on the dark 
energy equation of state parameter and its evolution by measuring the 
baryonic acoustic oscillations imprinted when matter and radiation 
decoupled, the distance-luminosity relation of cosmological 
supernovae, the evolution of the cosmic expansion rate, the growth 
rate of cosmic large-scale structure, and high-$z$ galaxy clusters.
The datasets from the \SPACE~ mission will represent a long 
lasting legacy for the whole astronomical community whose data will be
mined for many years to come.  


\end{abstract}

\section{Introduction and Scientific Background}

Our view of the Universe has changed dramatically over the past two 
decades through measurements of the cosmic microwave background (CMB), 
the large-scale structure of the local Universe ($z<0.3$) and the 
brightness of distant supernovae. Fully 96\% of the constituents consist of 
non-luminous and unidentified dark energy (73\%) and dark matter (23\%) 
that govern the expansion history and evolution of cosmic structure and 
leave their imprints on the structure and distribution of visible galaxies. 
These dark components are unexplained in standard physical theory, but
are, nevertheless, considered a natural feature of standard cosmology 
because they can explain a wide range of observations assuming only that 
dark energy produces a pressure countering gravity and dark matter behaves 
like ordinary matter in its effect on space-time. 

The acceleration of the expansion of the Universe is considered one of the 
most important discoveries of cosmology (\cite{Riess+98}, \cite{Perlmutter+99})
. Both of the extant explanations for accelerating space-time require new 
physics: a negative pressure component dubbed dark energy, or a modification 
of the law of gravity and, therefore, the standard framework underpinning 
cosmology. The first possibility, dark energy, is currently favoured but, 
nevertheless, provides little help because we have no plausible candidates 
from elementary particle theories (\cite{Padmanabhan07}). The simplest case 
is represented by the so called cosmological constant, $\Lambda$. 
Unfortunately, there is no theoretical justification for the size of the 
cosmological constant as inferred from the rate of universal acceleration. 
Arguments based on the standard model of 
particle physics yield values between $10^{50}$ and $10^{123}$ times larger 
than the observed value. The cosmological constant is only one of several 
candidates to explain the acceleration of the universe. Most possibilities 
can be parameterized by their equation of state, the ratio of the pressure 
exerted by the dark energy to the energy density of the field: 
$w = P/\rho c^2$. A cosmological constant necessarily implies $w = -1$. 
The simplest alternative quantum-field explanations allow $w$ to differ from 
$-1$ and to vary with time. Translating from linear time to observable 
redshift, $z$, we can parameterize $w(z) = w_0 + w_a(1-a)$, $a=1/(1+z)$. 
The magnitude of $w$ determines the rate of acceleration (through the 
associated pressure), so it can be characterized by measuring the rate of 
acceleration as a function of time for the history of the Universe, or 
equivalently, the Hubble parameter as a function of redshift, $H(z)$. 
Alternatively, the acceleration could be the result of a modified form of 
gravity, such as gravity that is a function of scale (e.g. 
\cite{Capozziello+07}), or a modification of the standard model through 
string or brane theories (e.g. \cite{Peacock+06}). Distinguishing between 
dark energy, modified gravity and different variants necessarily requires 
high-precision measurements of the cosmic expansion rate history $H(z)$, 
but the remaining differences among theories mean that even a very accurate 
$H(z)$ is inadequate to uniquely isolate one theory (e.g. \cite{Tegmark03}; 
\cite{Trotta06}; \cite{Linder07a}). 
The degeneracy occurs because a dark energy model and a modified
gravity theory can give identical cosmic expansion histories H(z), but
would give very different growth histories of cosmic large scale
structure (\cite{Lue+04}). However, the predicted growth rate of large-scale
structures varies among theories, and a measurement of structure
evolution can isolate the correct one. The best current constraints on 
$w$ from observations of distant supernovae are consistent with a cosmological 
constant: $-1 < w < -0.85$ with $10\%-20\%$ uncertainties assuming a flat 
Universe (\cite{Wood-Vasey+07}; \cite{Riess+07}; \cite{Sanchez+06}). 

The cosmological models make testable predictions about how the dark 
components affect a wide range of observable features that can be 
discriminated with high precision only using newly developed technology 
on spacecraft. In this framework, we designed a new space mission aimed 
at addressing the key questions of modern cosmology. This study led to 
a proposal that was submitted to ESA in response to the first call 
{\it Cosmic Vision 2015-2025}: the {\sl SP}ectroscopic {\sl A}ll-sky 
{\sl C}osmic {\sl E}xplorer 
(\SPACE~\footnote{http://urania.bo.astro.it/cimatti/space}). 
To solve the mystery of dark energy, \SPACE~ will determine $H(z)$ and 
constrain $w_0$ and $w_a$  by combining the observations of the Baryonic 
Acoustic Oscillations (BAO), the growth rate of structures, distant Type 
Ia Supernovae and high redshift galaxy clusters. But \SPACE~ will also
address several other key questions of modern cosmology besides dark
energy. The main scientific objectives of \SPACE~ are presented in the
next sections.

\section{\SPACE~ and the Power Spectrum of Density Fluctuations}

Thanks to the enormous volume sampled and the availability of
spectroscopic redshifts, \SPACE~ will be able to place constraints at an
unprecedented level on the power spectrum of matter fluctuations, $P(k)$,
which is a key input to theoretical models of structure formation.  The
spectrum of CMB temperature fluctuations, $C_l$, and the matter $P(k)$ are
subject to different parameter degeneracies. While CMB spectra depend on
the combination $\Omega_m h^{2}$, the matter $P(k)$ is sensitive to
$\Omega_m h$. Thus, combining these two datasets help to break
degeneracies and adds tremendous scope to tighten the limits on parameters
which result from the use of just one dataset in isolation. The $C_l$
spectrum expected from the \Planck~\footnote{http://www.rssd.esa.int/Planck} 
satellite will probe to smaller angular scales 
than WMAP\footnote{http://map.gsfc.nasa.gov/}, significantly increasing the 
overlap with the scales probed by the
SPACE measurment of $P(k)$. The massive volume covered by SPACE means that
we will be able to make the first compelling measurement of the turnover
in the matter $P(k)$. This feature is a prediction of the standard
paradigm of structure formation, imprinted when matter first started to
dictate the expansion of the Universe around 27,000 years after the Big
Bang. This measurement of the large scale power will be the first direct
probe of primordial fluctuations, which have only been measured to date
through ripples in the CMB. The position and shape of the turnover in
$P(k)$ depends upon the matter density of the universe, and how much of
this mass is in the form of baryons and massive neutrinos. By measuring
the slope of the power spectrum on scales larger than the turnover we will
be able to distinguish between different models of inflation, extending
the grasp of SPACE on the history of the universe right back to within a
tiny fraction second after the Big Bang.

\section{\SPACE~ and  Baryonic Acoustic Oscillations}

Baryonic acoustic oscillations (BAO) are small amplitude (5-10\%) 
modulations in the distribution of matter imprinted at the epoch when 
matter and radiation decoupled ($z\sim1000$). These ripples correspond to 
sound waves in the primordial photon-baryon fluid. The length scale of the 
oscillations is closely related to the sound horizon at decoupling, 
a scale ($\sim150$ Mpc) which is accurately known from measurements of 
CMB temperature fluctuations (\cite{Spergel+07}). Acoustic patterns imprinted 
on the cosmic microwave background have provided the most precise 
measurements of cosmological parameters by WMAP and form the basis for 
the even more precise measurements anticipated for the \Planck~.  
The physical scale of these oscillations will have 
different apparent sizes on the sky at different redshifts and in different 
cosmologies. Assuming that the distribution of galaxies reflects the 
distribution of all matter, this physical scale is measured from the 
spatial correlation (i.e. the power spectrum) of galaxies. The acoustic 
oscillation physical scale length is effectively a standard ruler whose 
apparent size can be measured in the transverse and the radial directions 
(3D) from the distribution of galaxies. Combining the apparent size with 
the known physical size gives the distance at the redshift of observation and 
makes it possible to determine the expansion rate as a function of redshift, 
$H(z)$ (\cite{Blake+Glazebrook03}; \cite{Hu+Haiman03}; 
\cite{Seo+Eisenstein03}; \cite{Wang06}). This technique is considered 
a very promising new method proposed to measure the evolution of the 
expansion rate and therefore the relative effects of dark energy and 
dark matter. The application of BAOs to extant galaxy surveys (in 
combination with other datasets or with restrictive assumptions like a 
flat Universe) has already yielded constraints on $w$ with uncertainties of 
only 10\% (e.g. \cite{Eisenstein+05}; \cite{Hutsi05}, \cite{Hutsi06}; 
\cite{Tegmark+06}; \cite{Percival+07}) even though this technique is 
only a few years old. The astrophysical effects which may cause deviations 
of the ideal BAO signature, for example nonlinear evolution of the density 
field, dynamical redshift-space distortions, galaxy biasing are now well 
understood from numerical calculations (\cite{Seo+Eisenstein03}, 
\cite{Seo+Eisenstein05}; \cite{Angulo+05}, \cite{Angulo+07}; 
\cite{Eisenstein+06}; \cite{Huff+06}). Angulo and collaborators 
(\cite{Angulo+07}) demonstrated that the primary limitation on the accuracy 
of the BAO method is sample variance. Thus, to minimize the uncertainty in 
$w$, the observations need simply to maximize the survey volume.  
To maximize the sample volume, \SPACE~  will perform a spectroscopic 
All-Sky Survey to measure galaxies over $3/4$ of the entire sky and over 
the redshift interval $0<z<2$, centered on the redshift $z\sim1$ at which 
the dark energy first starts to dominate the expansion of the Universe. 
The sensitivity requirement is $SNR\sim 3$ per resolution element at 
$H_{\rm AB}\sim23$. Taking spectra at near infrared wavelengths ensures that 
\SPACE~ covers the entire redshift range efficiently. The use of a 
space-based multi-object spectrometer makes it possible to get accurate 
redshifts of millions of galaxies over the whole sky in a short time. 
Fig. 1-2 show the power of the \SPACE~ BAO approach and the expected accuracy. 
It is clear that \SPACE~ will dramatically improve our knowledge of $w_0$ and 
$w_a$. Recent works emphasized the need to trace the evolution of dark
energy, $w(z)$, with high accuracy in order to derive the most stringent 
constraints on the dark energy (see \cite{Linder07a}). 

Thanks to the huge volume and wide redshift
range covered, \SPACE~ will achieve 0.5\% accuracy in the BAO scale 
measurement redshift slices with $\Delta z\sim0.5$. Even more accurate 
constraints on dark energy will result from a combination of independent 
and complementary datasets (e.g. see \cite{KetchumWang}). But even in the 
case of \Planck~ + \SPACE~ only (see Fig. 2), the constraints on $w$ will 
be $\approx$5 (\SPACE~ BAOs), $\approx$10 (\SPACE~ BAOs + current SNe
Ia), and $\geq$20 (\SPACE~ BAOs + \SPACE~ SNe) times more accurate than the 
current accuracy. This also highlights the "self--sufficiency" of \SPACE.

\begin{figure}\sidecaption
\resizebox{0.6\hsize}{!}{\includegraphics*{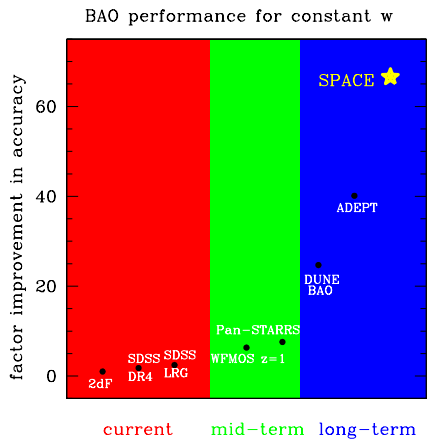}}
\caption{The relative accuracy with which various BAO experiments are 
forecast to measure the cosmological constant (or a fixed equation of 
state for the dark energy, i.e. $\Delta w = \Delta w_0$ and $w_a=0$), 
without combining with any other cosmological constraints. SPACE will 
yield a BAO measurement that is over 65 times more accurate than can 
be made currently with the 2dF survey, and outperforms any other 
proposed BAO experiment. The relative accuracy is computed by taking 
the square root of the ratio of the effective survey volume to the 
volume of the 2dF.}
\label{fig:baocomp_baugh}       % Give a unique label
\end{figure}

In 2006, NASA established the {\it Dark Energy Task Force} (DETF\footnote{
http ://home.fnal.gov/$\sim$rocky/DETF/} with the purpose of studying and
defining the accuracies reachable with future experiments dedicated to 
dark energy. 

None of the currently planned ground-based BAO experiments can compete 
with the accuracy on $w$ reachable with a SPACE-like survey. For instance,
the SDSS-III BOSS (Baryon Oscillation Spectroscopic Survey) experiment 
based on measuring BAOs from Luminous Red Galaxies at $z<0.7$ over a 
10,000 deg$^2$ sky area (see http://www.sdss3.org/cosmology.php) would 
achieve a DETF Figure of Merit (FoM) about an order of magnitude less than 
a SPACE-like survey to $H_{\rm AB}$=23 over 20,000 deg$^2$ and
\SPACE~ fully meets the requirements of the DETF 
which recommended that the goal for new survey designs should be that 
their {\it Figure of Merit improves by a factor of three between 
different Stages of dark energy missions} (\cite{Albrecht+06}).  
Combining the \SPACE~ BAO measurement with a {\it long-term Stage IV 
experiments}, such as LSST, JDEM and SKA, produces a further factor of 
three gain in the constraints on dark energy. Thus, \SPACE~ falls in the 
next category of {\it Stage V} experiments capable to improve our 
understanding of the dominant systematic effects in dark energy 
measurements and, wherever possible, reduce them. 

\begin{figure}\sidecaption
\resizebox{0.6\hsize}{!}{\includegraphics*{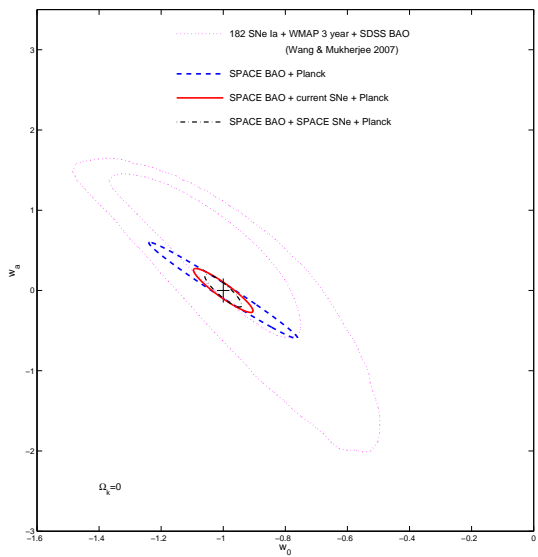}}
\caption{The accuracy on dark energy parameters reachable by combining 
\Planck~ and \SPACE. The ellipses indicate the joint 68\% confidence 
interval constraints on the dark energy equation of state parameter 
($w_0$) and its evolution with redshift ($w_a$). The dotted lines refer 
to the combination of three-year WMAP data (\cite{Spergel+07}), 182 SNe Ia
(\cite{Riess+07}, including \cite{astier} data), and SDSS BAO measurements
(\cite{Eisenstein+05}) (see \cite{Wang07b} for more details).
}
\label{fig:w0wa_space_cur}       % Give a unique label
\end{figure}

\section{\SPACE~ and the Growth Rate of Cosmic Structures}

The growth rate of cosmic structures due to gravitational instability 
offers the capability to distinguish between different theories that 
produce the same $H(z)$ and provides an independent and novel probe of 
the nature of dark energy or possible modifications to gravity. 
The growth rate of density fluctuations at a given redshift is
$f_g = d ln (D) / d ln (a)$, where $D(a)$ (or $D(t)$) is the growing 
solution of the equation describing the time evolution of density 
fluctuations (which depends on the gravity theory and on the expansion 
history). It has been shown that to a good approximation $f$ is well 
described as $f_g(z) \sim \Omega_m (z)^{\gamma}$ (\cite{linder05}) for a 
wide range of models, including modifications of General Relativity.
$f_g(z)$ can be measured from the impact of gravitationally induced peculiar 
motions (i.e. motions of galaxies in their local rest frame relative to 
the systemic flow due to the Hubble expansion) on the pattern of galaxy 
clustering. These motions produce an anisotropy in the galaxy redshift-space 
correlation function, $\xi(r_p,\pi)$, in directions parallel and perpendicular 
to the line of sight. The coherent bulk flows towards overdensities lead to 
an enhancement of the clustering signal on large scales, quantified as a 
compression in the iso-correlation levels of $\xi(r_p,\pi)$. The compression 
is directly proportional to the value of $f_g(z)$, modulo the {\sl bias 
factor} of the objects used to trace the matter distribution, $b$, measured 
by the parameter $\beta= f_g(z )/b$ (e.g. \cite{Hamilton+01}). 
The feasibility of this technique at intermediate redshifts has just been 
demonstrated in an important breakthrough using the VVDS redshift survey and 
by measuring $\beta$ with 40\% uncertainty from an area of 4~deg$^2$ using 
6000 galaxy redshifts over a wide redshift range (\cite{Guzzo08}; 
\cite{Pierleoni+07}). Other recent studies emphasized the power of this 
approach for differentiating between dark energy and modified gravity 
(\cite{Linder07b}, \cite{Wang07}). \SPACE~ will make a dramatic improvement 
over this, thanks to the enormous volume sampled, even in slices of 
$\Delta z=0.2$,  and to the deep infrared selection ($H_{\rm AB}<23$), 
that allows 
the $N(z)$ to extend significantly above $z=1.5$, with a mean density 
always larger than $\sim0.02$~gal~h$^3$Mpc$^{-3}$ to $z\sim 2$ even with a 
sampling of $1/3$ galaxies. Our simulations clearly show that \SPACE~ will 
be unique in providing a precision measurement of the growth history of 
cosmic structures, reaching 0.5\% accuracy in $\beta$ in redshift slices 
of $\Delta z=0.2$, and measuring directly the growth function $f_g(z)$ to 
1\% accuracy (Fig. 3). The bias factor ($b$) will 
be extracted in each redshift slice using higher-order galaxy clustering 
(\cite{Verde+01}) or via self-consistent density-reconstruction methods 
(e.g. \cite{Sigad+00}).

\begin{figure}\sidecaption
\resizebox{0.6\hsize}{!}{\includegraphics*{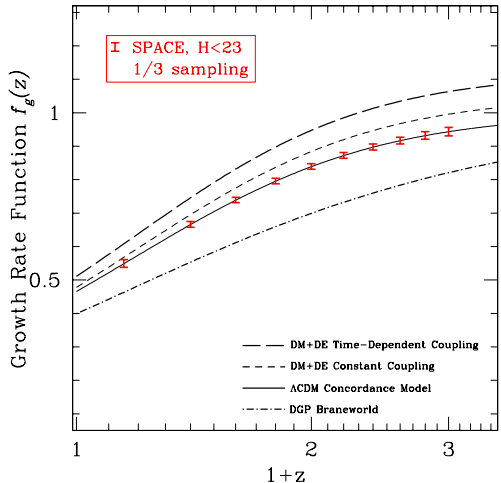}}
\caption{The predicted performance of \SPACE~ in measuring the growth rate 
of density fluctuations as a function of redshift, $f_g(z)$ using the 
anisotropy of galaxy correlations, assuming a $\Lambda$CDM fiducial model 
(see \cite{Guzzo08}).  \SPACE~ will disentangle to high accuracy the 
cosmological constant theory from variants of dark energy (here two possible 
cases are shown - \cite{Amendola00}). The combination of the fully 
independent measurements of $H(z)$ from BAOs and $f_g(z)$ from redshift 
distortions will provide a direct test of whether cosmic acceleration is 
due to a modification of Einstein theory of gravitation or to new physics 
beyond the standard model.}
\label{fig:Growthfunction}       % Give a unique label
\end{figure}

\section{\SPACE~ and distant Type Ia Supernovae}

Using Type Ia Supernovae (SNe Ia) as standard candles to measure $H(z)$
is the method by which the accelerating expansion of the Universe was 
first discovered (\cite{Riess+98}; \cite{Perlmutter+99}). \SPACE~ 
observational strategy can be designed to allow the discovery of SNe Ia 
and to obtain their spectroscopic redshifts out to $z\sim2$ very 
efficiently thanks to the simultaneous multi-object spectroscopy over a wide 
field. For instance, the repeated observation of a  4 deg$^2$ Deep Field 
($H_{\rm AB}<26$) would allow the identification of  $\sim2300$ SNe to 
$z\sim2$ in  
about 5--7 months spread over 1 year, where 1 visit would take 4 days to 
cover 4 deg$^2$ to $H_{\rm AB} \sim26$,  and each visit would be repeated every 
7--10 days. We also note that near-IR spectroscopy is advantageous because 
less affected by dust extinction, and hence very useful for high-z SNe.
The details of the \SPACE~-SNe program will depend on the developments of 
other future projects in space and on the ground (e.g. SNAP, Pan-STARRS, 
LSST, ...). One could imagine a scenario where SNAP and \SPACE~ 
will be operational in the same time frame in order to provide the strong 
synergy with SNAP discovering the SNe through imaging and \SPACE~ obtaining 
their spectra and redshifts.

\section{\SPACE~ and high-redshift galaxy clusters}

Galaxy clusters are the end products of the hierarchical build-up of 
small systems into increasingly larger structures up to $10^{15}$~M$_\odot$.  
Since clusters arise from rare high peaks of primordial density perturbations, 
their number densities and mass distribution, i.e. their mass function 
$n(M,z)$, is highly sensitive to the matter density parameter $\Omega_m$, 
and dark energy parameters ($\Lambda, w$) which control the rate at which 
structure grows, as well as to the normalization of the matter power spectrum, 
$\sigma_8$. At $z>1$, the existing surveys do not yet place strong 
constraints on dark energy due to the small volumes probed and because only 
a handful of systems have been identified to date (e.g. \cite{Rosati+04}, 
\cite{Mullis+05}, \cite{Stanford+06}). \SPACE, with its unmatched 
spectroscopic speed in near-IR, will provide spectroscopic confirmation of 
all the thousands of clusters at $z>1$ detected in the next generation 
near-IR, Sunyaev-Zeldovich (SZ) and X-ray large area surveys (e.g. SPT in 
2007+, Planck in 2008+, eROSITA in 2012+).  \SPACE~ will identify bona-fide 
virialized structures, the ones to be compared to the theoretical cluster 
mass function, and it will resolve the cases of contamination from 
point-like radio sources and AGN that are expected to plague future SZ and 
X-ray searches for distant clusters. Thus, \SPACE~ will unleash the full 
potential of the next generation cluster surveys by allowing the best 
possible knowledge of systematic errors, with small statistical uncertainties 
limited only by the volume of the observable Universe. We recall that 
constraints from cluster surveys generally exhibit parameter degeneracies 
different from those of other techniques, such as CMB, Type-Ia SNe and BAO 
described above. The pursuit of these complementary means, based on different 
physical properties of the Universe, is the most powerful way to unveil the 
nature of the dark energy and dark matter. In addition, \SPACE~ itself, 
thanks to the unbiased near-infrared selection and the spectroscopic survey 
will locate for the first time several tens of thousands of clusters directly 
in three dimensions out to $z\sim2.5$ and over a large mass range. 

\section{From Dark Matter to Baryons: galaxy formation and evolution}

In the standard model of galaxy formation, galaxies coalesce from 
gas that cools inside dark matter halos (\cite{White+Rees78,baugh06}). 
Determining the physical processes that regulate the growth of galaxies and 
their link with massive black holes is one of the outstanding problems of 
modern astrophysics.  By observing galaxies down to faint magnitudes and 
sampling {\it randomly} one of every three of these faint galaxies, \SPACE~ 
will measure the characteristics of more than half a billion galaxies as a 
function of 
their environment at sensitivities impossible to obtain from ground-based 
telescopes.  We take this opportunity to underline why \SPACE~ must 
observe in the near-infrared. The strong optical-ultraviolet spectral 
features of galaxies well below $L_*$ that are primary redshift and 
diagnostic tracers fall, in the redshift range $z\sim0.5-3$, in the near-IR. 
Their list includes: Balmer and D4000 continuum breaks, strong H Balmer 
lines, CaII H\&K lines, as well as the well-studied diagnostic emission lines 
such as [O II]$\lambda$3727, Balmer lines, [O III]$\lambda \lambda$4959,5007, 
and several high ionization lines useful for identifying the presence of an 
active galactic nucleus.  At higher redshifts, \SPACE~ will obtain redshifts 
and diagnostics of the young stellar populations and the interstellar medium 
using rest-frame ultraviolet lines. \SPACE~'s combination of wavelength 
coverage, resolution and sensitivity will remove the degeneracy limiting the 
photometric SED fitting (reddening vs. red stars). Examples of what 
\SPACE~ will be able to accomplish include: (1) the evolution of the 
distribution functions with the highest accuracy possible and limited only 
by cosmic variance (e.g. luminosity, stellar mass) significantly below the 
characteristic mass, $M_*$, (2) the cosmic evolution of red/passive galaxies 
at $0<z<2$ and up to their formation epoch, (3) to investigate how the 
properties of galaxies depend on the density of their surroundings, (4) to 
determine the merger rate as a function of redshift, (5) to determine black 
hole masses from line widths and compare them to the stellar content, mass, 
and age of the host galaxy, to directly probe feedback processes in the 
largest ever high-z sample of AGN, (6) to determine the causes of "downsizing", whereby star-formation histories correlate with stellar mass, and the active sites of star-formation move to increasingly higher mass galaxies with redshift, (7) to provide a unique resource for measuring all aspects of the spatial distribution of galaxies and clusters on smaller scales, including the higher order correlation functions (i.e. the amplitudes of the N-point correlation functions) which are sensitive to the growth of structure through gravitational instability and offer a direct route to measuring galaxy bias (see section 2.2). Moreover, \SPACE~ will enable new facilities such as LOFAR, eROSITA, WISE, SKA to reach their full capability by removing the significant "bottleneck" in obtaining the redshifts for the large survey samples.

\section{Galaxies and QSOs in the early Universe}

In the area of galaxy and AGN co-evolution, \SPACE~ will provide a unique contribution through a spectroscopic Deep Survey, by targeting galaxies at redshifts as high as 10, when the Universe was a few hundred million years old. The \SPACE~ Deep Survey will be designed to study large samples at  high redshifts following a strategy of this type: (1) $Z$-, $J$- and $H$-band imaging of a 10 deg$^2$  area to $H=26$, $J\sim28$, $Z\sim28$, (2) narrow-band imaging of the same area, (3) spectroscopy to $H=25$ over the same area and with a sampling rate of 90\% (mostly at $2<z<7$), (4) the candidates at $z>7$ will be pre-selected from broad-band colors (\cite{Bouwens+04}, \cite{Bouwens+06}) using the \SPACE~ imaging data themselves and will be repeatedly observed for secure spectroscopic identification. Based on the observed galaxy luminosity function at $z=6$ (\cite{Bouwens+06}), extrapolated at higher redshift under the assumption of simple luminosity evolution, in an area of 10~deg$^2$, there will be $\sim800$ galaxies at $z>7$ with $H<26^m$, $\sim500$ at $7<z<8$, $\sim200$ at $8<z<9$, and a few tens above $z=9$. As the most massive objects at these redshifts are extremely rare, the large FOV of \SPACE~ compared to JWST gives it an enormous advantage: only $\sim0.3$ massive galaxies should be present in the JWST field of view of $\sim10$ arcmin$^2$. \SPACE~ is the natural complement to JWST for the study of high redshift galaxies. With narrow-band filters, it will be possible to search for Ly-$\alpha$ emitters too faint to be detected in the broad-band \SPACE~ images by selecting objects showing an excess of flux in narrow-band filters with respect to the broad-band one. For instance, filters centered at $\sim1.03$ and $\sim1.34~\mu$m, allowing an extensive Ly-$\alpha$ emitter search at $z\sim7.7$ and $z\sim10$. At $AB=26$, \SPACE~ should find  $\sim1200$ ($\sim30$) emitters at $z\sim7.7$ ($z\sim10$) objects with Ly-$\alpha$ fluxes as low as $\sim4\times10^{-18}$~erg~s$^{-1}$cm$^{-2}$ with $S/N\sim10$.  A fraction (up to 10\%) of these sources is expected to be powered by very massive, metal-free stars, the so called Population III stars (\cite{Scannapieco+03}). These objects can be selected from \SPACE~ data owing to their very large Ly-$\alpha$ equivalent width and He II $\lambda$1640 emission (\cite{Shaerer03}).  Finally, \SPACE~ will also obtain the definitive measurement and spectrum of the so far elusive cosmic Near Infrared Background (NIRB) (e.g. \cite{Kashlinsky05} and \cite{thompson07} and references therein). Sources at $z>5$ contributing to the reionization of the Universe are expected to produce a strong feature in the NIRB, with a well defined maximum at $\sim1\mu$m with $\nu I_\nu\sim1$~nW~m$^2$sr$^{-1}$  and a strong break below $1~\mu$m due to the strong absorption by the intergalactic medium (IGM) below rest-frame 912~\AA (\cite{Salvaterra+06}; 
\cite{Choudhury+Ferrara06}). 

The enormous sky coverage of the \SPACE~ all-sky survey will also reveal several hundred QSOs at $z>6$ (including a few tens of objects at $z>9$) and shed light on the evolution of supermassive black holes. 
To estimate the detected number of high-redshift QSOs, a $z=6$
luminosity function was constructed using current observational
constraints (\cite{Fan+04}, \cite{Willott+05}), assuming 2/3 of the 
sky is covered, and 1/3 sampling rate within this sky coverage. 
Three scenarios were considered, in order to model the evolution of 
the $z\geq6$ QSO LF: no evolution from the $z=6$ LF, linear evolution, 
and Eddington (exponentially decaying) evolution with an $e$-folding 
time of $\tau=45$ Myr. These assumptions yield a detection rate of 
approximately 200, 150 and $\le 1$ (70, 30, $<1$) QSOs at $z=7$ ($z>9$).
The \SPACE~ spectral resolution is good enough to study the absorption of photons from Ly$\alpha$ and Ly$\beta$ transitions through the IGM as a function of redshift. The masses of QSOs can be measured from the \SPACE~ spectra using the C IV $\lambda1549$~\AA~ line (\cite{Vestergaard02}). Currently, it takes about 8 hours with a 8m telescope to obtain a suitable near-infrared spectrum of a QSO at $z\sim6$ (e.g. \cite{Simcoe06}), whereas in 15 minutes \SPACE~ would go 3 magnitudes fainter. Before 2017, programs on ground-based telescopes will search for high-redshift QSOs by imaging in several optical and near-infrared bands, and using drop-out techniques. However, such techniques are limited by (a) the near-infrared imaging depth achievable from the ground for a large sky area, (b) the limited redshift ranges available to the drop-out technique, (c) the large fraction ($>95\%$) of drop out candidates that are cool stars, and (d) the OH sky lines covering $>40\%$ of the near-infrared spectrum. Ground-based surveys are expected to find only a relatively small number of high-redshift QSOs (for a recent example of the process, see \cite{Venemans+07}). For a significant increase in the number of objects, and to really open up the unexplored redshift range ($z>7$), we need to search for high-redshift QSOs over the entire available all sky with spectra that cover a large wavelength range free from by atmospheric emission lines.

\section{The Near-Infrared view of our Milky Way Galaxy}
\label{sec:5}
At low galactic latitudes source crowding represents a problem for multi-slit spectroscopic surveys. However, the extreme versatility of the instrument we have envisioned allows us to perform integral field spectroscopy using coded masks (Hadamard transforms). A \SPACE~ Galactic-plane survey will be feasible with a relatively modest investment of time by covering in integral field mode the strip within $\pm0.5^\circ$ centered on the Galactic Equator between $\pm60^\circ$ of galactic longitude to magnitude $AB\sim20$. Other  examples of key contributions of \SPACE~ in Galactic studies include the census of  ultracool dwarfs in the vicinity of the Sun and the estimate of the lower mass function in star-forming regions.

\section{\SPACE~ vs ground-based observations}

All the scientific goals described above need the coverage of the largest
possible volume of the Universe, and are reachable within reasonable 
timescales only if the observations are done from space and in the near-IR
by taking advantage of the sky background lower by a factor of 
$\approx$500 than from the ground, and the lack of OH emission lines 
and telluric absorptions. From the ground, 30\% of the wavelength range 
between 1 and 2~$\mu$m is invisible, and for the remaining fraction the 
bright and variable night sky lines dominate (at the $\sim95\%$ level !) 
the background. At low spectral resolution ($R\sim500$), their filling 
factor is virtually 100\%. Even at $R\sim2000$, less than half the near-IR
spectral range is free from night sky line emission (\cite{Martini00}). 
These emission lines, variable in intensity on time scales of minutes, 
add background noise, subtraction residuals and scattering that make 
spectral line and redshift determination of faint sources from the 
ground nearly impossible, no matter how long one integrates. 
The \SPACE~ uninterrupted spectral coverage of $\sim$0.8--1.8$\mu$m coupled 
with the near-IR selection of the all-sky survey sample ($H_{\rm AB} <23$) 
ensures a
survey speed, a redshift coverage of $0<z<2$ and a spectroscopic redshift 
measurement success rate that would be impossible to obtain from 
the ground. This huge advantage opens also the possibility to perform 
a \SPACE~ deep and wide-field ($\approx$10 deg$^2$) spectroscopy survey 
down to limiting magnitudes ($H_{\rm AB} <25-26$) (see Section 4)
that would be completely unreachable
with ground-based observations even in the era of extremely large 
telescopes ($>$10m diameter).

\section{Simulations}
\label{sec:6}

In order to assess the mission feasibility and efficiency, and to define the most suitable characteristics of the instrumentation, the performances of \SPACE~ were extensively simulated both for imaging and spectroscopy. 

\subsection{Spectroscopy}
\label{sec:6.1}

In the case of spectroscopy, we explored different instrument configurations 
using the VVDS slit positioning software (\cite{Bottini+05}) and using both 
real and artificial (from the Millennium Simulations) galaxy catalogs with 
magnitudes matched to the sensitivity limits of the All-Sky and Deep Survey. 
The software does not allow the overlap of spectra both in the spatial and 
in the wavelength directions (Fig. 4). 
The mission efficiency was evaluated as a function of instrumental parameters 
and galaxy properties, conservatively allowing 2 empty pixels between 
spectra to accommodate spectral distortions.  The best results are 
obtained with: 4k$\times$4k detector, pixel size $0.375''$, 15 \AA/pix 
dispersion, 2 pixels per DMD along dispersion, spectrum length of 670 pixel, 
spectral resolution of 400. This combination allows a sampling of about
1 of every 3 galaxies to $H_{\rm AB} =23$(AB), giving a $\approx$33\% efficiency. 

\begin{figure}
%\includegraphics[width=1.0\textwidth]{simulations.eps}
\includegraphics[width=0.98\textwidth]{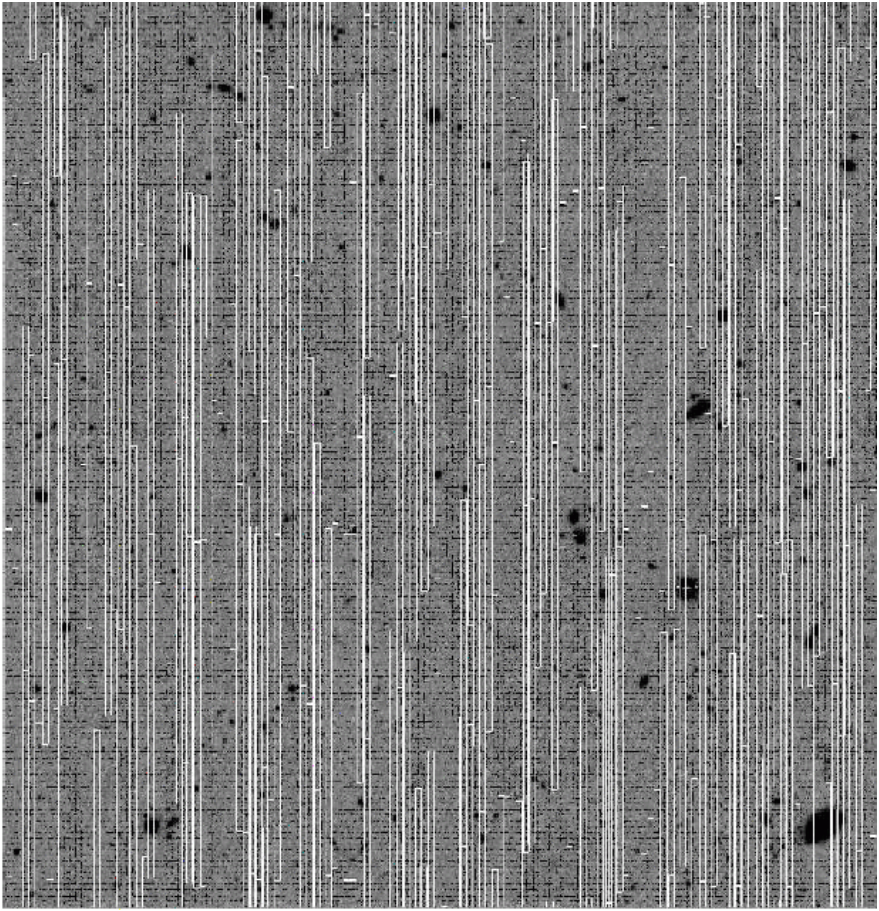}
\caption{A simulated $5'\times7'$ field showing the location of the spectra of the selected targets.}
\label{fig:simulations_ED}       % Give a unique label
\end{figure}
\begin{figure}
%\includegraphics[width=1.0\textwidth]{spectrum.eps}
\includegraphics[width=0.97\textwidth]{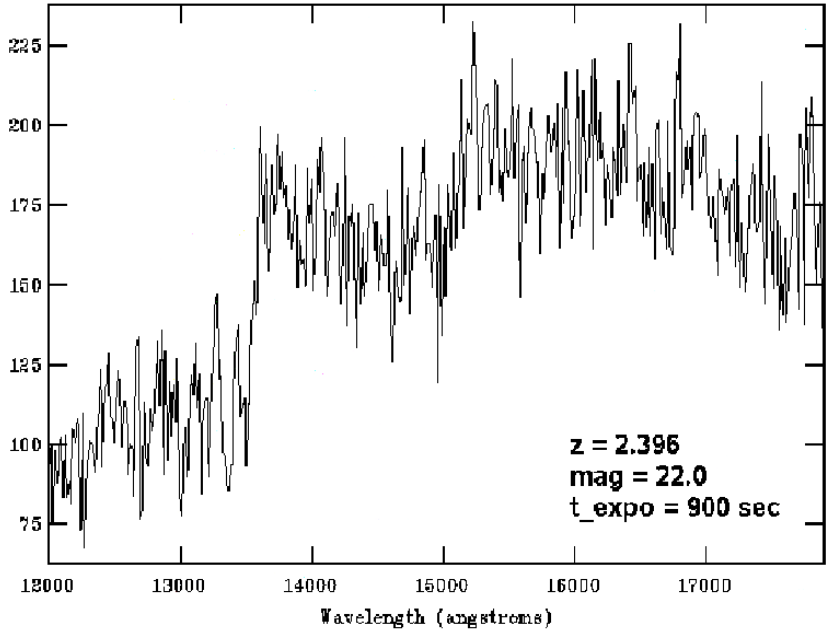}
\caption{Simulated spectrum of an early-type passively evolving galaxy at $z>2$. The D4000 break and the main absorption lines (CaII H\&K) are clearly detected. It is impossible to obtain such a spectrum with ground-based near-IR spectroscopy due to the sky background, OH line contamination and telluric absorptions.}
\label{fig:spectrum_ED2}       % Give a unique label
\end{figure}

With the above parameters, we simulated spectra including background and detector noise. Template spectra (\cite{Kinney96}, \cite{Mannucci01}, \cite{Shapley03}) of a variety of galaxy types (early-type, star-forming, starbursts, AGN, Lyman-break, etc.) provided the input to the simulations as a function of the magnitude and redshift for a total of about 180,000 spectra at $0.2<z<2.4$ for the all-sky survey ($H_{\rm AB} <23$) and about 1,000 at $2.4<z<10$ for the Deep Survey ($H_{\rm AB} <25-26$). The results show that the dispersion and the resolution of \SPACE~ are adequate to reliably identify the main emission and absorption features of all galaxy types (Figure~\ref{fig:spectrum_ED2}). Redshifts have been automatically measured with EZ (the software tool used in VVDS and zCosmos projects) and other tools specifically developed for the simulations. The low background, wide wavelength coverage, and moderate spectral resolution produce a success rate of spectroscopic redshift measurements between 
80\%--99\%  for the planned integration times of 900 sec (All-Sky Survey) and 7 hours (Deep Survey) for all galaxy types and redshifts. The typical accuracy on individual redshifts $\sigma_z=0.001$. In comparison, with the same integration time of 900 sec, slitless spectroscopy would provide a S/N ratio about 10 times lower.  

\subsection{Imaging}
\label{sec:6.2} 

We verified that the photometric catalogues derived from \SPACE~ imaging will provide high quality target lists for the spectroscopic observations. We used the NICMOS H-band observations of the Hubble Ultra Deep Fields (HUDF) with PSF $FWHM\sim0.15''$ to match closely the PSF delivered by the 1.5m \SPACE~ mirror and resampled the data at the \SPACE~ scale of $0.375''$/pixel. The vast majority of sources detected in this image, reaching well beyond $H_{\rm AB} =27$, are still easily seen in the resampled image without blending or confusion. We measured blending and confusion as a function of magnitude using source detection software for the original and resampled HUDF images. The results are: (1) at the $H_{\rm
AB}<23$ (all sky survey), $97\%$ of the sources are still individually detected in the resampled image, (2) at $H_{\rm AB}=26$, only $\sim15\%$ of the sources appear blended with neighbouring sources, (3) source counts from the original and resampled image are consistent well beyond $H_{\rm AB}=26$.

\section{The \SPACE~ survey programmes}
\label{sec:7}

The main objective of \SPACE~ will be to execute a Core Science Program in about 3.5 years. However, given the power and flexibility of  \SPACE~ in many fields of astrophysics, it will allow also for Guest Observer Science Programs. 
The Core Program will be composed of:

\begin{itemize}
\item
The \SPACE~ all-sky survey will observe the entire sky at galactic latitudes $>\pm18^\circ$ ($\sim70\%$ of the total). This corresponds approximately to 28,500 square degrees and 71,000 satellite pointing (1 pointing = 0.4 deg$^2$). Assuming 20 minutes per pointing and an observing efficiency of 75\%, this will require 2.7 years. The expected number of galaxy spectra with a 1/3 target random sampling is (50,000/3)$\times71,000\times0.4\sim0.5\times10^9$. The selection of the spectroscopic targets will be done taking an acquisition image in the H-band at $AB\sim23.5$ immediately before the spectroscopic observation. Targets will be selected using on-board software applied to the images taken with \SPACE~ itself
, following the approach currently adopted in ground-based spectroscopic surveys (e.g. VVDS, zCOSMOS). As a byproduct, the acquisition images will produce the deepest all-sky imaging survey at $|b|> \pm18^\circ$.

\item
The \SPACE~ Deep Survey will target a 10 deg$^2$ field (25 \SPACE~ pointings) down to $H_{\rm AB}=25$, with a target sampling rate of 90\%, and without any color pre-selection. About 200,000 objects (stars+galaxies) are expected in 1 deg$^2$ at $H_{\rm AB}<25$. With a multiplex of  $\sim6000$ objects and an integration time of about 7 hours per observation to reach a sufficient S/N, we need $\sim[(200,000 \times 0.4 \times 0.90)/6000] \times 25$~pointings $\times 7h \approx 5$~months. We recall that the $z>7$ galaxy candidates will be pre-selected using the \SPACE~ imaging in z,J,H bands and will be "compulsory" targets repeated in each of the 12 spectroscopic observations done for each of the 25 pointings so that they will accumulate a total integration time of (7~h$\times$12) = 84 hours (per pointing) needed to reach a sufficient S/N for J, H = 26
(AB). The time needed for the broad-band z-, J- , H-band and narrow-band imaging will be negligible with respect to the time dedicated to spectroscopy. With an appropriate strategy of repeated visits, the Deep Survey could be used also for detecting high-z SNe (see Section 2.3)

\item
The \SPACE~ Galactic-plane survey will cover in Integral field mode the strip within $\pm0.5^\circ$ centered on the Galactic Equator between $\pm60^\circ$ of galactic longitude. We will allocate 4~hr per pointing, i.e. 10 hours per square degree and 50 days total for the entire survey.
\end{itemize}

Allocating a fraction of the time to a Guest Observer Program will add value to the mission. We will split the Core Science program in a four phases lasting approximately 9 months each, allowing the GO observing time to be ramped up over time in a manner similar to the phasing in past missions such as ISO and Spitzer. This will allow the core science program to be completed in a timely manner while providing the community with early instrument performance information and science results to use in the planning of observations for other studies.

\section{\SPACE~ Mission concept}
\label{sec:8}

\subsection{Driving requirements}
\label{sec:8.1}
The science requirements identified in the previous sections place firm constraints on mission concept and the design of instrument. 

The first need is to access the whole celestial sphere for about 3 years. The core observing program requires $0.1''$ pointing stability for at least 20~min of typical integration time, high temperature stability and good access to the ground station for data downlink and telemetry. All these requirements are ideally met by a large-amplitude halo orbit around the second Earth-sun Lagrange point (L2). A halo orbit at L2 maximizes the availability of the sky for observation with respect to the visual and thermal disturbance from the Sun, Earth and Moon. The thermal environment is very stable and, with the exception of solar storms, the radiation environment is relatively low. On a halo orbit of sufficient amplitude eclipses can be avoided altogether, while the constant satellite-sun-Earth geometry eases the thermal control and the communications. 
L2 is considered the ideal vantage point for an IR observatory and, following
WMAP, several mission (e.g. Herschel, Planck, GAIA, JWST) will be 
operated at L2.

In order to measure a large number of sources, one needs multi-object spectroscopic capability over an area of the order of 1 square degree. In order to observe faint sources, one needs to operate in slit mode, exploiting the celestial background from space $\sim500$ times lower than from the ground. The sensitivity requirement also drives the telescope diameter, whereas the large field of view drive the effective focal length of the optical system. The broad wavelength range and resolution of the spectrograph drives the choice of materials (mirrors vs. lenses), the detector technology, the instrument temperature, and ultimately the multiplexing capability of the instrument. In particular, the need for extensive multi-object capability suggests considering Micro-Electro-Mechanical-Systems (MEMS), following the approach pioneered for NIRSPEC on JWST  (\cite{Moseley+00}, \cite{MacKenty+Stiavelli00}).  

Having set through performance simulations the basic requirements on the instrument parameters (see Sect. 11), 
we have evaluated several possible solutions and converged finally on an elegant design that satisfies all requirements. To keep complexity and cost to a minimum we aimed for a single-mode instrument rather an all-purpose machine with a variety of arms and configurations. Nevertheless, we came out with a solution which inherently provides an extraordinary degree of versatility without adding complexity, weight or cost to the basic hardware needed for our main survey program. Using MEMS for target selection we have several millions of randomly addressable optical switches. Depending on the MEMS configuration, and on the selection of a filter or a prism in a filter wheel (we may have a single filter wheel,
which would then represent the only moving part on \SPACE) \SPACE~ can perform imaging, slitless and slit spectroscopy (our main observing mode), and even wide-field integral field spectroscopy using Hadamard transforms. 

\subsection{Payload description}
\label{sec:8.2 }
The \SPACE~ payload is a compact and modular instrument, fitting below the primary mirror structure within a cylindrical envelope of 1.5~m diameter (the size of the primary) by 60~cm height. To accommodate for our large field of view, the beam is split into 4 identical channels, each one taking a $90^\circ$ sector of the cylinder.  
Each channel is a relatively conventional spectrograph with a MEMS device located at the intermediate focal plane. For the MEMS device we have opted for a Texas Instrument DMD of $2048\times1080$ micro-mirrors, whereas for the detector we use a 4k$\times$4k infrared Focal Plan Assembly similar to the one developed for JWST. 
Each micro-mirror images a square $0.75''\times0.75''$ field onto $2\times2$ detector pixels. Hereafter we will examine the baseline optical design and
the main component of a spectrograph more in detail. 

\begin{figure}
%\includegraphics[width=1.0\textwidth]{OTA.eps}
\includegraphics[angle=-90,width=1.0\textwidth]{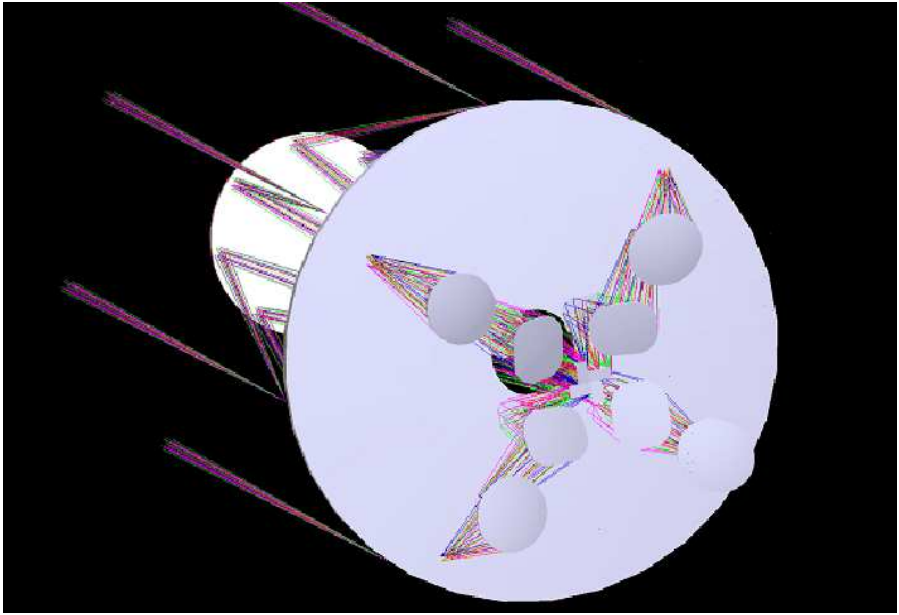}
\caption{Optical Telescope Assembly and fore-optics system (four channels).}
\label{fig:OTA_ED}       % Give a unique label
\end{figure}

\begin{enumerate}
\item
The telescope is a Ritchey-Chr\'etien F/5.83, with a 1.5m F/2.7 primary.  A Ritchey-Chr\'etien telescope, in which both mirrors are hyperbolic, is free from coma and allows for easy optical testing. 

\item
The telescope focal plane falls on a 4-face pyramid mirror providing a total coverage of $51'\times27'$ ($\sim0.4$ square degrees). The layout sketched in Figure~\ref{fig:pyramid} preserves the symmetry around the optical axis allowing the four channels to be identical with substantial cost saving in the design, fabrication and testing. It still allows to "tile" the sky with multiple contiguous observation to map large fields without significant losses in coverage. 

% I do this figure with a side caption, using the instructions on the usrguid3.tex file
\begin{figure}\sidecaption
   \resizebox{0.4\hsize}{!}{\includegraphics*{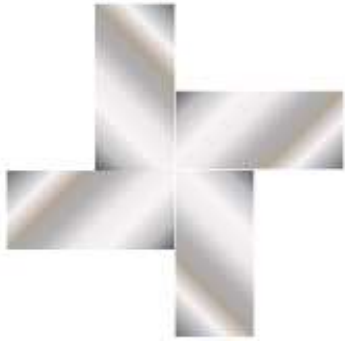}}
\caption{Four-fold symmetric cross for the SPACE~ pyramid mirror allowing for 4 identical optical channels.}
\label{fig:pyramid}       % Give a unique label
\end{figure}

\item
Each subfields of the pyramid mirror is reflected toward a fore-optics system, made of 4 mirrors (Figure~\ref{fig:foreoptics_ED}) contained in a layer $\sim400$~mm deep under the primary support structure. The fore-optics system projects on the DMD an undistorted image of the focal plane with a scale of $0.75''$/DMD facet and excellent optical quality. 

\begin{figure}
%\includegraphics[width=0.5\textwidth]{foreoptics_top.eps}
%\includegraphics[width=0.5\textwidth]{foreoptics.eps}
%\includegraphics[width=0.5\textwidth]{foreoptics_top_ED.eps}
%\includegraphics[width=0.5\textwidth]{foreoptics_ED2.eps}
\includegraphics[width=1.0\textwidth]{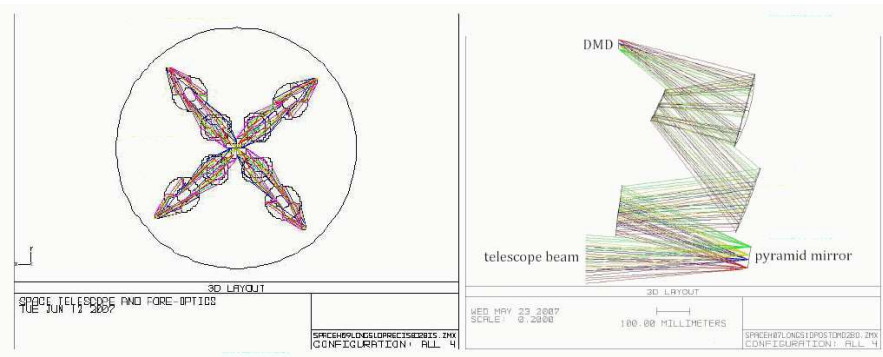}
\caption{Fore-optics system. Left: top view (all channels); Right: side view (one channel). }
\label{fig:foreoptics_ED}       % Give a unique label
\end{figure}

\item
The DMD micro-mirrors are $13.6~\mu$m on a side and tilt along their diagonal by $\pm 12^\circ$. When the DMD facets are tilted to their 'ON' position, the beam is reflected in a direction approximately normal to the DMD plane, while when they are turned 'OFF' they reflect at an oblique angle towards a beam dump. 

\item
The collimator collects the beam reflected by the DMD relaying an image at infinity. Its design is similar to that of the fore-optics, with 4 mirrors located in the same 400mm thick layer (Figure~\ref{fig:collimator_ED}). The collimator, however, develops inward along a cord rather than outward along a radius, therefore the incoming and outcoming beams do not cross. The collimator produces a 66~mm pupil located nearly on the same plane of the pyramid mirror.  

\begin{figure}
%\includegraphics[width=0.5\textwidth]{collimator_top.eps}
%\includegraphics[width=0.5\textwidth]{collimator.eps}
%\includegraphics[width=0.5\textwidth]{collimator_top_ED2.eps}
%\includegraphics[width=0.5\textwidth]{collimator_ED.ps}
\includegraphics[width=1.0\textwidth]{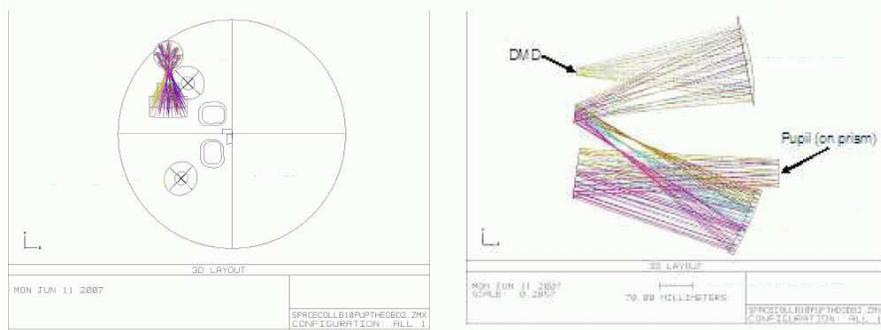}
\caption{Left: top view layout of the collimator (one channel). The location of the fore-optics mirrors of two channels (Figure~\ref{fig:foreoptics_ED}) is also shown to illustrate the different orientation of the collimator (along a cord) with respect to the the fore-optics (along a radius) after reflection on the DMD. Right: side view layout of the collimator.}
\label{fig:collimator_ED}       % Give a unique label
\end{figure}

\item
The prism is located in the immediate vicinity of the pupil. Having set at $R\sim400$ target, we have considered a number of solutions based on both grisms and prisms in different configurations and materials. Our baseline choice is for a single pass prism + folding mirror placed immediately after the pupil. For the imaging mode, the prism + folding mirror are substituted by a simple $45^\circ$ folding mirror and broad-band filter. A single filter wheel per channel allows to insert the other required imaging filters (e.g. z, J, narrow band). 

\item 
The camera reimages the DMD plane onto the detector plane with a 1:2 ratio. The camera has 4 lenses, including 2 aspheric surfaces, and is achromatic and diffraction limited between 0.6 and 1.8~$\mu$m. The length of a 0.8-1.8~$\mu$m $R\sim400$ spectrum is $\sim670$ pixels, slightly dependent on the field position.  Thanks to a folding mirror, the detector can be located in the area below the pyramid mirror. Having all IR detectors located close to each other simplifies the cooling system. 
\end{enumerate}

% I do this figure with a side caption, using the instructions on the usrguid3.tex file
\begin{figure}\sidecaption
%\resizebox{0.6\hsize}{!}{\includegraphics*{prism_camera.eps}}
\resizebox{0.6\hsize}{!}{\includegraphics*{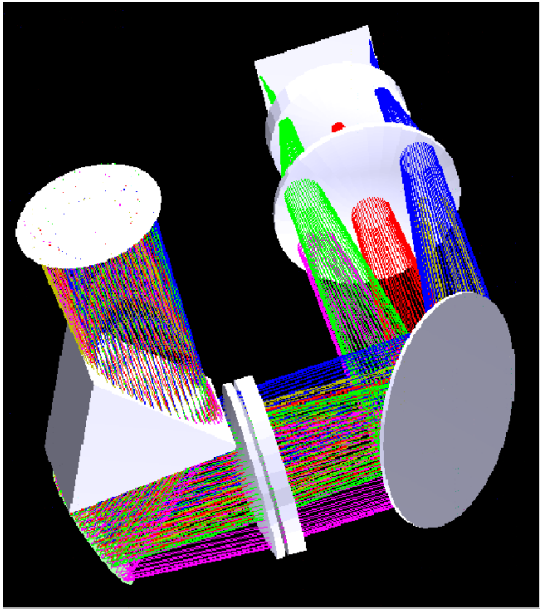}}
\caption{Prism and camera 3-d layout. A $45^\circ$ folding mirror can be inserted in the camera to relay the image in the most convenient position at center of the spacecraft.}
\label{fig:prism_camera_ED}       % Give a unique label
\end{figure}

\subsection{Detectors}
\label{sec:8.3}

\SPACE~ will utilize HgCdTe Focal Plane Arrays (FPAs). Teledyne devices, based on the Hawaii-2RG multiplexer, with long wavelength cut-off at $\sim1.8~\mu$m are ideally suited because of their modest sensitivity to the spacecraft thermal emission and to the low dark current levels exhibited at relatively high operating temperatures, $\sim140-150$~K, but more conventional 2.5~$\mu$m devices can also be used, since the spacecraft will be passively cooled. This technology is mature. The IR channel of the HST/WFC3 camera is equipped with  FPAs with 1.72~$\mu$m cutoff wavelength on a 1k$\times$1k multiplexer. 
The most recent WFC3 parts attain $\sim80\%$ QE nearly flat over a $1.0-1.70~\mu$m range, readout noise in double correlated sampling $<25$~e/read and dark current $\sim0.01$~e/s/pixel at 140~K. JWST devices for NIRCAM and NIRSPEC with $2.5~\mu$m cutoff have also been produced in 2k$\times$2k format, with excellent performance. Note that the substrate removal process, necessary to eliminate radiation induced glow, extends the spectral response down to $0.4~\mu$m, allowing an extension of the wavelength range down to e.g. $0.60~\mu$m that we are considering for \SPACE.

% I do this figure with a side caption, using the instructions on the usrguid3.tex file
\begin{figure}\sidecaption
   \resizebox{0.4\hsize}{!}{\includegraphics*{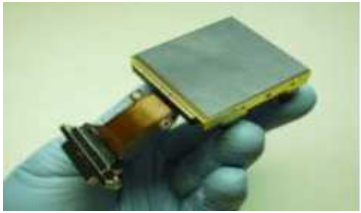}}
\caption{A 2k$\times$2k buttable Focal Plan Array developed for JWST.}
\label{fig:JWSTdetector}       % Give a unique label
\end{figure}

These devices can be butted together in a $2\times n$ configuration (Focal Plane Assembly, FPA). For \SPACE~ we are considering $2\times 2$ FPA similar to those developed for JWST, A 4k$\times$4k assembly allows to easily accomodate for the extension of the spectra.

\subsection{Digital Micromirror Devices (DMDs)}
\label{sec:8.4}

The innovative technical aspect of the \SPACE~ mission is the use of DMDs. We have taken a fresh look at these devices, that were originally considered for NIRSPEC on JWST, and concluded that they represent the ideal choice for \SPACE. 
Since their invention by L. Hornbeck at Texas Instruments (TI) in 1988, DMD have been serially produced with volumes exceeding 10 million units shipped by 2006. Today DMDs represent one of the leading technologies in digital imaging (e.g. DLP projectors). For this multi-billion dollar market huge investments have been made, reaching unique levels of quality and reliability. We intend to capitalize on these investments.

\begin{figure}
  \includegraphics[width=1.0\textwidth]{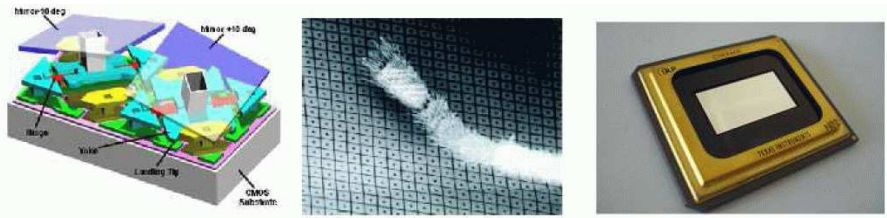}
\caption{Left: typical substructure of a TI DMD; center: DMD array with an ant leg for comparison; 3: packaged DMD CINEMA (2048$\times$1080) device.}
\label{fig:DMD_res}       % Give a unique label
\end{figure}

\subsubsection{What are DMDs?}
\label{sec:8.4.1}
DMDs are a particular type of Micro-Electro-Mechanical Systems (MEMS). MEMS integrate microscopic mechanical elements, in our case moving mirrors, together with their control electronics on a common silicon substrate. Whereas the control electronics is fabricated using standard integrated circuit processes, the mechanical components are fabricated using "micromachining" processes that selectively etch away parts of the silicon wafer or add new structural layers.
In the case of DMDs, the mechanical part is an array of up to 2.21 million aluminum micromirrors fabricated on top of a complementary metal oxide semiconductor (MOS) static random access memory (SRAM) array. Each micromirror, independently controlled, can switch along its diagonal thousands of times per second as a result of electrostatic attraction between the mirror structure and the underlying electrodes. The current generations of DMDs all exhibit a $+12^\circ /-12^\circ$ tilt angle, limited by a mechanical stop. Pixel pitch in currently available DMD is either $10.8~\mu$m ($0.6~\mu$m gap width) or $13.68~\mu$m ($0.68~\mu$m gap width) center-to-center. 
TI produces several types of DMDs in hermetic packages. Those with more than 2 million pixels are the $2048\times1080$ pixels "CINEMA" device (Figure 6.9.3) with $13.68\mu$m pixel pitch, and the $1920\times1080$ 1080p "High Definition TV" device with $10.6~\mu$m pixel pitch. For our baseline design we have considered the former, but optical solutions can be found for smaller devices. 

\subsubsection{Why DMDs? (instead of micro-shutter arrays)}
\label{sec:8.4.2}
The main alternative to DMD is the Micro-shutter array (MSA), which has been developed and optimally matched to the IRMOS/JWST. MSAs represent an outstanding technological achievement and are especially unique in their capability of operating at low temperatures. However, they come in much smaller size (at the moment $171\times365$ pixels) and are still prone to cosmetic defects. Other figures of merit (e.g. filling factor, contrast, reliability) do not appear to be superior to those achieved by the current generation of commercial DMDs.  DMDs are technologically mature, already in use in Astronomy (\cite{MacKenty+06}) and meet all the optical and technical needs of the \SPACE~ mission. The issues (optical quality, constrast, reliability, radiation environment, ....) related to the use of DMD in our mission are discussed in detail in the \SPACE~ proposal.
          
\section{Performance assessment with respect to science objectives}
\label{sec:9}
We have verified that with our proposed design \SPACE~ reaches the sensitivity goals stated in the Sect.~\ref{sec:6} and \ref{sec:7}. Using the baseline optical design described in the previous section we have estimated the signal to noise ratio per resolution element in 900 s of integration  (Fig. 13). A $SNR\sim 3$ at AB=23 is the sensitivity requirement derived in Section 3 for the All-sky survey. The sensitivity estimates for the Deep Survey are also compatible with the requirements. 

\begin{figure}
%\includegraphics[width=1.0\textwidth]{sensitivity.eps}
\includegraphics[width=1.0\textwidth,angle=360]{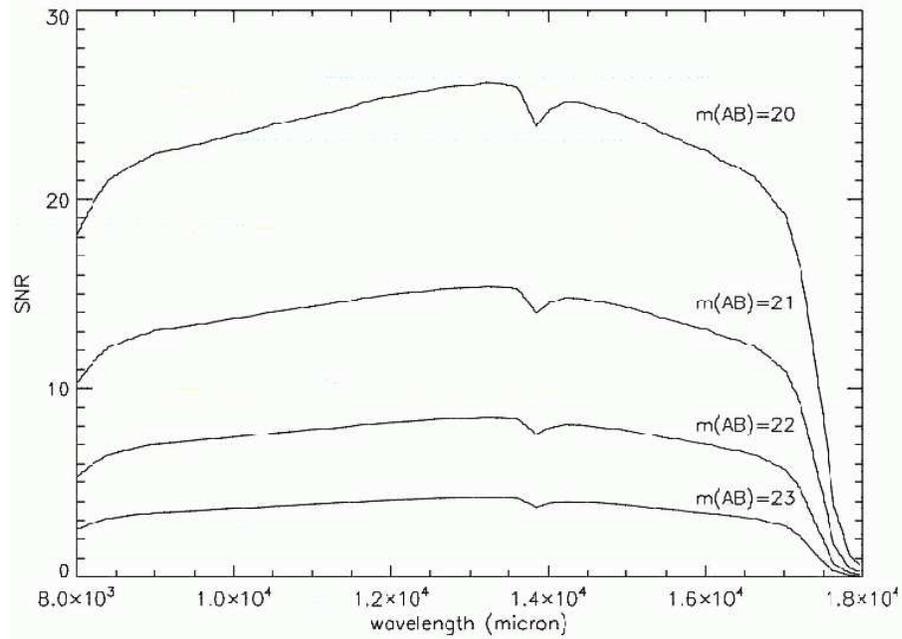}
\caption{Estimated SNR of SPACE~ spectra in 900s integration. We used detector parameters typical of WFC3/IR flight candidates. Transmission efficiencies of all reflective and refractive components are those of the WFC3/IR optical coating. Light losses due to the prism thickness are included, together with the most recent zodiacal background prediction for the SNAP mission also at L2.}
\label{fig:sensitivity_ED2}       % Give a unique label
\end{figure}

\section{Operating modes}
\label{sec:10}

\begin{figure}
%\includegraphics[width=1.0\textwidth]{ObservingStrategy.eps}
\includegraphics[angle=-90,width=1.0\textwidth]{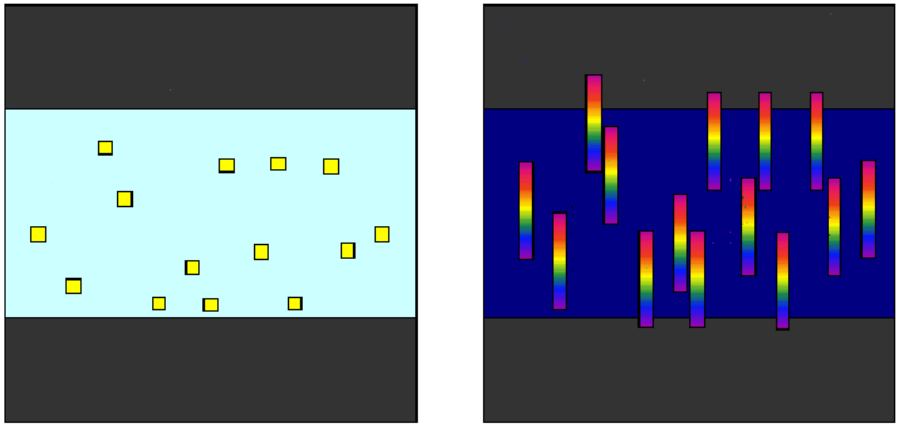}
\caption{Sketch of the data acquisition and observing procedure. Left: the DMD field is projected onto the detector (dark background) in broad band imaging mode. The light blue color indicates the high background, targets are represented by yellow squares. Right: all DMDs are turned off except those of the targets, the prism is inserted. The background is low (dark blue) and spectra are produced.}
\label{fig:ObservingStrategy_ED}       % Give a unique label
\end{figure}

The all-sky survey will be performed repeating the same basic observing mode, sketched in Figure~\ref{fig:ObservingStrategy_ED}. After locking the guider, a $\sim30$~s broad-band image will be taken and corrected for dark, bad pixels and flat field using calibration frames stored on-board. A software routine (similar to the one we used in Sect.~\ref{sec:6.1}) will then automatically select the targets, generate and upload the optimal DMD configuration. In the mean time, the prism will be inserted at the place of the broad-band filter. We allocate 5 minutes for these operations, which will proceed in parallel on the four channels. 

As mentioned in Sect.~\ref{sec:6}, refinements are possible, e.g. splitting the acquisition of spectra in multiple exposures to eliminate cosmic rays or to optimize the extraction of the bright sources. The HAWAII 2RG detectors allow to address and read individual pixels while the others integrate ("guide mode"). This capability may also be exploited to synchronize the readout of certain areas with the opening of the corresponding slits.

The long exposures performed for the Deep Survey will be more efficient, avoiding the frequent broad-band acquisition. We plan to add telescope dithering to recover spatial resolution and dead pixels. For the integral field spectroscopic survey of the Galactic plane, we will use Hadamard transform spectroscopy. Proposed by J. Mather for NIRSPEC on JWST and successfully validated with IRMOS (\cite{Fixsen+MacKenty07}), this technique can be regarded as a variation of long slit spectroscopy. By opening multiple slits at the same time, overlapped spectra are obtained. By using different slit patterns, the entire set of overlapped spectra can be eventually disentangled and the entire data cube ($x,y,\lambda$) reconstructed. Our estimates for the Galactic plane survey allocates four hours per pointing, which corresponds to individual exposures of the order of order of $\sim15$~s. The lower signal to noise of the raw images will allow for a more robust compression factor than the noiseless 5$\times$ used for the standard observations. The actual implementation of the Hadamard transform mode will require further analysis.

\section{Mission Operations and Data Analysis}
\label{sec:11}
Soyuz-Fregat is our proposed launcher. Soyuz, equipped with the Fregat upper stage, launched from Kourou, is capable of delivering a 2-ton spacecraft directly into L2 transfer orbit without any need for large on-board propulsion. A typical duration of the transfer is 4 months. Once L2 has been reached, the delta-v requirements for orbit maintenance and formation keeping are small, a few m/s/year.
The launcher provides a fairing 3.8 m wide and up to 8.5m high (with $14.5^\circ$ taper above 5~m), which can easily accommodate the spacecraft configuration with an external baffle for stray light rejection. 

In the survey phase, the sequence of operations is particularly simple (20' to 30' pointing, small slew, new pointing, and so on), easily automated and managed. Targeted observations will require similar sequences. A ground station such as Cebreros or New Norcia will have L2 above the horizon for more than 8 hours per day on average, with a minimum of 6 hours/day in the most unfavorable season for the given location in the Northern or Southern hemisphere. Our most conservative assumptions on compression (5~Mbit/s) and transmission (1.5~Mbit/s) require a station scheduled 3.5~hours per day for \SPACE.    

\SPACE~ telemetry and command will be received/transmitted by a dedicated antenna at a single ESA Ground Station having the task of gathering the downlinked housekeeping and science telemetry and uploading the commands needed to control spacecraft and payload. The Ground Station will be shared with other missions having similar visibility requirements. The telemetry will be sent through a dedicated redundant line of appropriate bandwidth to a Mission Operations Centre (MOC) under the control of ESA. The MOC is in charge of checking the housekeeping telemetry to verify the safety of spacecraft and payload, taking if needed the appropriate corrective actions by means of commands sent following pre-specified procedures. 
The stream of housekeeping and science telemetry, through an ESA-provided dedicated redundant line of appropriate bandwidth, will be then ingested into a Science Operations Centre (SOC) under the control of the \SPACE~ Consortium.  The SOC will also generate the commands needed for the satellite pointing and observations, and send them to the MOC in the form of agreed-upon procedures.

The SOC will distribute the data to a Science Data Centre (SDC), which
will focus on the spectra contained in the images, their extraction, analysis, distribution, archiving and generation of scientific results. The SDC will represent the interfaces between the project and the scientific community. The SDC will receives both the raw data and pre-reduced data from the SOC, adding all services needed to efficiently exploit their science content.

\paragraph{Acknowledgments} 
The authors aknowledge the support from the University of Bologna and 
the Space Telescope Science Institute. Thales Alenia - Milano (Italy)
is warmly acknowledged for its contributions and support.


\begin{thebibliography}{}
%\bibitem{RefB}Abdalla et al. 2007, arXiv:0705.1437; 
\bibitem{Albrecht+06}Albrecht A., et al. 2006, DETF report, http://home.fnal.gov/$\sim$rocky/DETF/; 
\bibitem{Aldering}Aldering, G. 2001, LBNL report number LBNL-51157.
\bibitem{Amendola00}Amendola, L., 2000, Phys.Rev. D, 62, 3511 ; 
\bibitem{Angulo+05}Angulo R., et al. 2005, MNRAS, 362, L25 ; 
\bibitem{Angulo+07}Angulo R., et al. 2007, astro-ph/0702543; 
\bibitem{astier}Astier P. et al. 2006, A\&A, 447, 31 
\bibitem{baugh06}Baugh C.M. 2006, Reports on Progress in Physics, 69, 3101 
%\bibitem{RefB}Begelman, MBHC., Volonteri, MBH, Rees, MBHJ., 2006, MNRAS,370,289; 
\bibitem{Blake+Glazebrook03}Blake  \& Glazebrook K. 2003, ApJ, 594, 665; 
\bibitem{Bottini+05}Bottini D. Et al. 2005, PASP, 117, 996; 
\bibitem{Bouwens+06}Bouwens  R.J., Illingworth G.D., Blakeslee J.P. Franx M., 2006, ApJ, 653, 53; 
\bibitem{Bouwens+04}Bouwens R. et al. 2004, ApJ, 616, L79; Bouwens R. et al. 2004, ApJ, 624, L5; 
\bibitem{Capozziello+07}Capozziello et al. 2007, arXiv0706.1146; 
\bibitem{Choudhury+Ferrara06}Choudhury T.R. \& Ferrara A., 2006, MNRAS, 371 L55; 
%\bibitem{RefB}Cowie L. Et al. 1996, AJ, 112, 839; 
%\bibitem{RefB}Dow-Hygelund C.C. et al., 2006, ApJ, 660, 47; 
\bibitem{Eisenstein+05}Eisenstein D. et al. 2005, ApJ, 633, 560; 
\bibitem{Eisenstein+06}Eisenstein D. et al. 2006, astro.ph.4362E; 
%\bibitem{RefB}Fan, X., Carilli, C. L., Keating, B., 2006, ARA\&A, 44,415; 
%\bibitem{RefB}Fan, X., et al., 2001, AJ, 122, 2833; 
\bibitem{Fan+04}Fan, X., et al., 2004, AJ, 128, 515; 
\bibitem{Fixsen+MacKenty07}Fixsen \& MacKenty 2007, in prep.; 
\bibitem{Guzzo08}Guzzo L. et al. 2008, Nature, in press 
\bibitem{Hamilton+01}Hamilton A.J.S. 2001, Technical Report, NASA GSFC; 
%\bibitem{RefB}Hirata C.M. et al. 2007, arXiv:astro-ph/0701671; 
%\bibitem{RefB}Hoekstra H., et al. 2006, ApJ, 647, 116; 
%\bibitem{RefB}Horton A. et al., 2004, SPIE, 5492, 1022; 
\bibitem{Hu+Haiman03}Hu \& Haiman 2003, PhRvD, 68, 6; 
\bibitem{Huff+06}Huff E. et al. 2007, APh, 26, 351 ; 
\bibitem{Hutsi05}H\"utsi G. 2005, A\&A, 449, 891; 
\bibitem{Hutsi06}H\"utsi G. 2006, A\&A, 459, 375; 
%\bibitem{RefB}Jarvis et al, 2006, ApJ, 644, 71; 
\bibitem{Kashlinsky05}Kashlinsky A., 2005, PhR, 409, 361; 
\bibitem{Kinney96}Kinney, A., Calzetti D., Bohlin R.C. et al. 1996, ApJ, 467, 38 
\bibitem{linder05}Linder E. 2005, Physical Review D, vol. 72, Issue 4, id. 043529  
\bibitem{Linder07a}Linder E., arXiv:0708.0024 
\bibitem{Linder07b}Linder E., arXiv:0709.1113 
\bibitem{Lue+04}Lue A., Scoccimarro R., Starkman G.D. 2004, Phys.Rev. D, 69, 124015; 
\bibitem{MacKenty+Stiavelli00}MacKenty \& Stiavelli 2000, ASP Conf. 195, 443; 
\bibitem{MacKenty+06}MacKenty et al. 2006, SPIE 6269, 37M; 
%\bibitem{RefB}Madau, P., 2007, 2007, astro-ph/0701394; 
%\bibitem{RefB}Majumdar S. and Mohr J.J. 2003, ApJ, 585, 603 ; 
%\bibitem{RefB}Majumdar S. and Mohr J.J. 2004, ApJ, 613, 41; 
\bibitem{Mannucci01}Mannucci, F., Basile F., Poggianti B. 2001, MNRAS, 326, 745 
\bibitem{Martini00}Martini P. \& DePoy D.L., 2000, Proc. SPIE, Vol. 4008, p. 695; 
%\bibitem{RefB}Matsumoto et al., 2005, ApJ, 626, 31; 
%\bibitem{RefB}Mattila K., 2006, MNRAS, 372, 1253; 
%\bibitem{RefB}McLure, R., Dunlop, J., 2004, MNRAS, 352, 1390; 
\bibitem{Moseley+00}Moseley et al. 2000, ASP Conf. 207, 262; 
\bibitem{Mullis+05}Mullis C. et al. 2005, ApJ, 623, L85; 
\bibitem{Padmanabhan07}Padmanabhan, T. 2007, arXiv0705.2533; 
\bibitem{Peacock+06}Peacock, J.A., et al 2006, ESA-ESO Working Group Report on Fundamental Cosmology, astro-ph/0610906; 
\bibitem{Percival+07}Percival et al. 2007, arXiv0705.3323; 
\bibitem{Perlmutter+99}Perlmutter, S., et al. 1999, ApJ, 517, 565; 
%\bibitem{RefB}Phillips, M.M. 1993, ApJ, 413, L105; 
\bibitem{Pierleoni+07}Pierleoni M. et al. 2007, submitted ; 
%\bibitem{RefB}Réfrégier A. et al. 2006 arXiv:astro-ph/0610062 ; 
\bibitem{Riess+98}Riess, A.G., et al. 1998, AJ, 116, 1009; 
\bibitem{Riess+07}Riess, A.G., et al. 2007, ApJ, 659, 98; 
\bibitem{Rosati+04}Rosati P. et al. 2004, AJ, 127, 230; 
%\bibitem{RefB}Rosati et al. 02, Ann.Rev.Astron. Astrophys, 2002, 40, 539; 
\bibitem{Salvaterra+06}Salvaterra R., Magliocchetti M., Ferrara A., Schneider R., 2006, MNRAS, 368, L6; 
\bibitem{Sanchez+06}Sanchez, A.G., et al. 2006, MNRAS, 366, 189 ; 
\bibitem{Scannapieco+03}Scannapieco E., Schneider R., Ferrara A., 2003, ApJ, 589, 35; 
\bibitem{Shaerer03}Schaerer D., 2003, A\&A, 397, 527; 
\bibitem{Shapley03}Shapley, A., Steidel C.C., Pettini M. et al. 2003, ApJ, 588, 65
%\bibitem{RefB}Schuecker P, Bohringer H, Guzzo L, Collins CA, Neumann DM, et al. 2001, Astron.  Astrophys. 368, 86; 
%\bibitem{RefB}Sembolini E., et al. 2006, A\&A 452, 5;
\bibitem{Seo+Eisenstein03}Seo H.-J. \& D. Eisenstein 2003, ApJ, 598, 720; 
\bibitem{Seo+Eisenstein05}Seo H.-J. \& D. Eisenstein 2005, ApJ, 633, 575; 
\bibitem{Sigad+00}Sigad Y. et al. 2000, ApJ, 540, 62; 
\bibitem{Simcoe06}Simcoe, R.A., 2006, AJ, 653, 977; 
\bibitem{Spergel+07}Spergel et al. 2007, ApJS, 170, 377; 
%\bibitem{RefB}Springel V. Et al. 2005, Nature, 435, 629; 
\bibitem{Stanford+06}Stanford S.A. et al. 2006, ApJ, 646, L13; 
%\bibitem{RefB}Stark D. P., Loeb A., Ellis R.S., 2007, astro-ph/0701882; 
%\bibitem{RefB}Sullivan, M, et al., 2006, ApJ, 648, 868; 
\bibitem{Tegmark03}Tegmark, M. 2003, Science, 296, 1427; 
\bibitem{Tegmark+06}Tegmark et al 2006, PhRvD, 74, 12; 
\bibitem{thompson07}Thompson R. et al 2007, ApJ, 657, 669 
\bibitem{Trotta06}Trotta R., 2006, arXiv:astro-ph/0607496
%\bibitem{RefB}Troxel, M.A.; Branch, D.; 
%\bibitem{RefB}Garnavich, P.; 
%\bibitem{RefB}Baron, E.; Jeffery, D.; 
\bibitem{KetchumWang}Ketchum, W., and Wang, Y. 2007, submitted to ApJ (2007); 
\bibitem{Venemans+07}Venemans, B., et al., 2007, MNRAS, 376L, 76
\bibitem{Verde+01}Verde L. et al. 2001, MNRAS, 325, 412; 
\bibitem{Vestergaard02}Vestergaard, M. 2002 ApJ, 571, 733-752
%\bibitem{RefB}Voit GM, 2005, Rev Mod Phys.  77, 207; 
\bibitem{Wang06}Wang Y. 2006, ApJ, 647, 1; 
\bibitem{Wang07}Wang Y. 2007, arXiv:0710.3885 
\bibitem{Wang07b}Wang Y., Mukherjee P. 2007, PRD, in press (arXiv:astro-ph/0703780 
\bibitem{White+Rees78}White S.D.M., Rees M. 1978, MNRAS, 183, 341; 
%\bibitem{RefB}White SDM, Navarro JF, Evrard AE, Frenk CS. 1993, Nature 366, 429; 
%\bibitem{RefB}White, R.L., et al., 2003, AJ, 126,1 ; 
\bibitem{Willott+05}Willott, et al., 2005, ApJ, 633, 630; 
\bibitem{Wood-Vasey+07}Wood-Vasey, W.M., et al. 2007, astro-ph/0701041;
\end{thebibliography}


\end{document}